\documentclass[times, twoside, watermark]{zHenriquesLab-StyleBioRxiv}
\usepackage{blindtext}

\leadauthor{K. M. Kabusure, P. Piskunen}

\title{Controlling Raman enhancement in particle-aperture hybrid nanostructures by interlayer spacing}
\shorttitle{Preprint}

\author[1,2,$\dagger$]{Kabusure M. Kabusure}
\affil[1]{Department of Physics and Mathematics, University of Eastern Finland, Yliopistokatu 2, P.O. Box 111, 80101, Joensuu, Finland}
\affil[2]{Department of Chemistry, University of Eastern Finland, Yliopistokatu 7, P.O Box 111, 80101, Joensuu, Finland}

\author[3,$\dagger$]{Petteri Piskunen}
\affil[2]{Biohybrid Materials, Department of Bioproducts and Biosystems, Aalto University, P.O. Box 16100, 00076, Aalto, Finland}

\author[2]{Jarkko J. Saarinen}

\author[3,4,\Letter]{Veikko Linko}

\affil[4]{Institute of Technology, University of Tartu, Nooruse 1, 50411, Tartu, Estonia}

\author[1,\Letter]{Tommi K. Hakala}

\affil[$\dagger$]{These authors contributed equally}

\begin{document}

\maketitle

\begin{abstract}
Here we show how surface-enhanced Raman spectroscopy (SERS) features can be fine-tuned in optically active substrates made of layered materials. To demonstrate this, we used DNA-assisted lithography (DALI) to create substrates with silver bowtie nanoparticle-aperture pairs and then coated the samples with rhodamine 6G (R6G) molecules. By varying the spacing between the aperture and particle layer, we were able to control the strength of the interlayer coupling between the plasmon resonances of the apertures and those of the underlying bowtie particles. The changes in the resulting field enhancements were confirmed by recording the Raman spectra of R6G from the substrates, and the experimental findings were supported with finite difference time domain (FDTD) simulations including reflection/extinction and near-field profiles. 
\end {abstract}


\begin{corrauthor}
veikko.pentti.linko@ut.ee /
tommi.hakala@uef.fi

\end{corrauthor}

\section*{Introduction}

Metallic nanostructures support surface plasmon resonances with characteristic high near field enhancements that can be utilized, for instance, in establishing strong coupling between plasmons and nano-emitters, lasing, and sensing.\cite{bellessa2004strong, hakala2017lasing, piliarik2009surface, azzam2020ten, ma2014explosives} Plasmonic resonators can be coupled by placing two or more nanostructures into close proximity to each other, resulting in a further enhancement of the near fields and significant resonance wavelength shifts due to normal mode coupling.\cite{PhysRevB.72.165409} Plasmonic nanostructures are typically fabricated using common solid-state lithographic techniques, but during the past decade, self-assembled DNA nanostructures have been widely employed as an alternative route to create nanophotonic systems.\cite{tan2011building, shen2018dna, pilo-pais2017sculpting, Kuzyk2018nanophotonics, Heuer-Jungemann2021engineering} Especially, modular DNA origami objects\cite{rothemund2006folding, douglas2009self} have been increasingly utilized as templates in developing plasmonic nanoparticle assemblies, optically active substrates, single-molecule tracking and sensing devices, nanoantennas, metasurfaces, and photonic crystals.\cite{kuzyk2012dna, gopinath2016engineering, funck2018sensing, gopinath2021absolute, trofymchuk2023gold, sanz2023dna, mostafa2024bulk, kanehira2024watching, sikeler2024dna, posnjak2024diamond} Versatile DNA nanostructures have also been used in lithographic settings as masks in pattern transfer\cite{jin2013metallized} and in fabricating accurate inorganic\cite{surwade2011molecular, surwade2013nanoscale, Piskunen2021Biotemplated, diagne2016dna, thomas2020dna, shen2021three, mao2023molecular} and metallic nanoshapes\cite{Piskunen2021Biotemplated, shen2015custom, shen2018plasmonic} with intriguing optical properties.\cite{shen2018plasmonic, kabusure2022optical, kabusure2023raman}

\begin{figure}[ht]
    \centering
    \includegraphics[width=0.95\columnwidth]{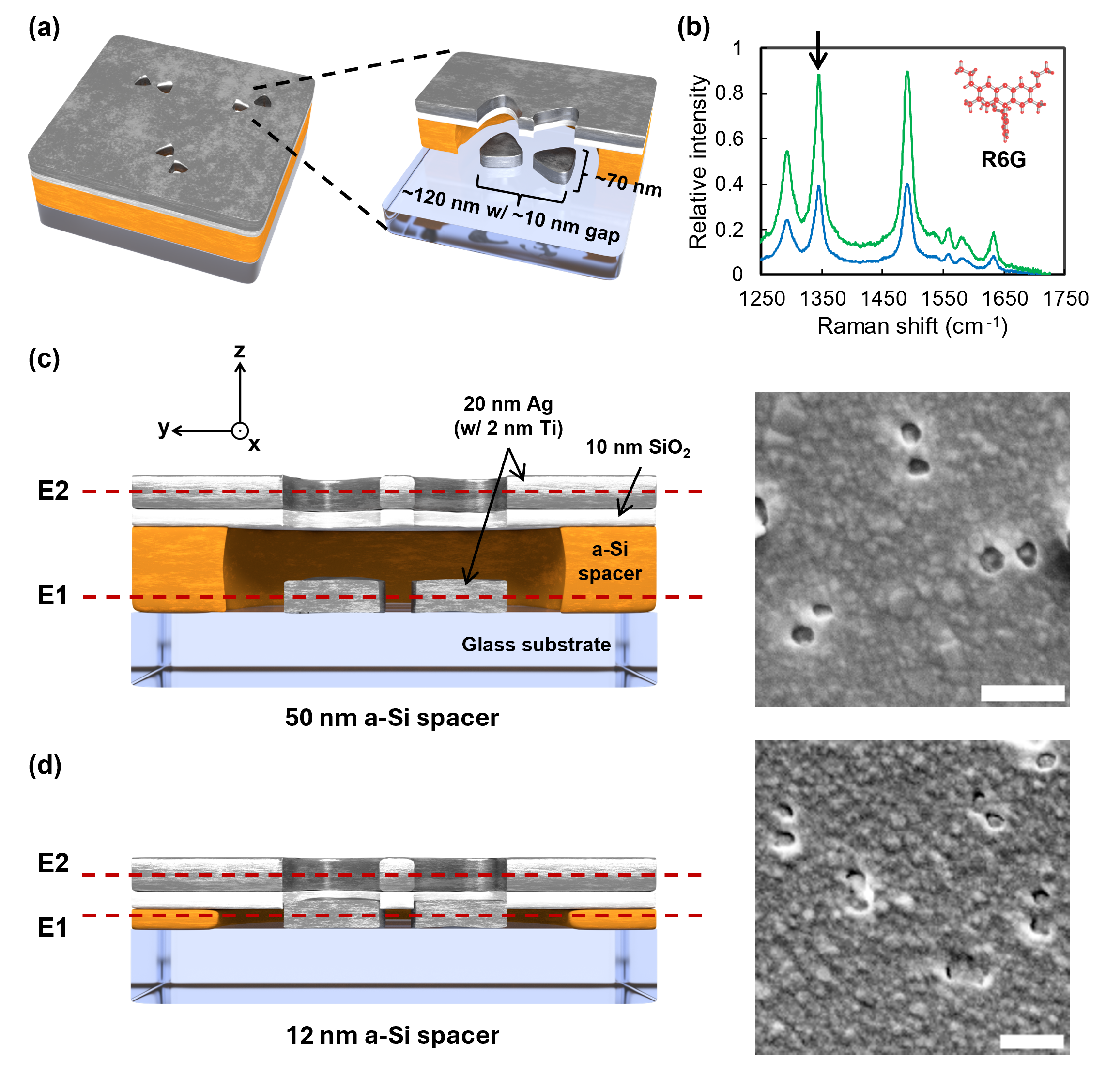}
    \caption{(a) Schematic view of the DALI-fabricated substrate showing the layered geometry and typical lateral dimensions of the aligned particle and aperture. (b) Schematic Raman spectra of rhodamine 6G (R6G, inset)-coated hybrid nanostructures. The intensity of the Raman signal can be controlled by tuning the interlayer spacing between the particles and apertures (here a green sample shows stronger interlayer coupling than the blue one at the Raman relevant wavelength range). The arrow indicates the Raman shift of interest (at 877 nm with Raman excitation of 785 nm). (c-d) Schematic cross-sectional view and the SEM image of the ready substrate with 50 nm or 12 nm a-Si spacer layer, respectively. Dashed lines (E1 and E2) indicate the horizontal planes at which the FDTD simulations have been carried out. The coordination system corresponds to the one used in Fig.~\ref{FDTDE1E3}. The scale bars in the SEM images are 200~nm.}
    \label{Schematic}
\end{figure}

In this article we use a DNA-assisted lithography (DALI) -based strategy \cite{shen2018plasmonic, shen2018dna-assisted} to create optically active substrates with aligned bowtie-shaped silver particle-aperture pairs in a layered configuration of materials (Fig.~\ref{Schematic}a). The substrate with hybrid nanostructures is covered with rhodamine 6G (R6G) molecules that serve as Raman-active reporters of the achieved field enhancement (Fig.~\ref{Schematic}b). The optical substrates are similar to what we have presented recently,\cite{kabusure2023raman} but here we systematically vary the interlayer spacing between the aperture and particle layers (Fig.~\ref{Schematic}c,d).

We hypothesize that since both the particles and apertures have plasmonic properties, the strength of the interlayer coupling between their resonances could be fine-tuned by varying their distance, i.e., the thickness of an a-Si spacer film that separates the layers. Furthermore, the changes in the interlayer coupling strength could then be observed as differences in R6G Raman intensities as schematically shown in Fig.~\ref{Schematic}b. To test and verify this hypothesis, we used four different spacer layer thicknesses and measured the Raman intensities from the R6G-coated substrates. We also performed detailed finite difference time domain (FDTD) simulations that fully supported the experimentally obtained results. The results show that it is indeed possible to find an optimal interlayer spacing that maximizes the field enhancement within the wavelength range of interest, i.e., within the range of the Raman excitation laser and the most prominent Raman transitions of R6G. We therefore believe that by optimizing the geometry of such DALI-fabricated hybrid nanostructures, it is possible to create custom, highly efficient optically active substrates with fully tailored surface-enhanced Raman spectroscopy (SERS)\cite{hering2008sers} features.

\section*{Methods}

\subsection*{Fabrication of particle-aperture nanostructures}

The substrates containing particle-aperture pairs were fabricated similarly as before.\cite{kabusure2023raman} The fabrication scheme is based on the DALI method, \cite{shen2018plasmonic, shen2018dna-assisted} where DNA origami nanostructures are used as templates/masks for pattern generation. Briefly, the process starts with growing an a-Si layer (Fig.~\ref{Schematic}, yellow layer) on top of a glass substrate. Here, the thickness of the deposited a-Si layer was varied between 12 and 50 nm (thicknesses of 12, 32, 42, or 50 nm were selected) to modulate the spacing between active layers (Fig.~\ref{Schematic}c,d). Then a 10-nm thick SiO$_2$ layer (Fig.~\ref{Schematic}, white layer) was selectively grown onto the a-Si surface with deposited DNA origami bowties, thus creating origami-shaped openings in the SiO$_2$ layer. The grown SiO$_2$ layer was used as a mask for etching the underlying a-Si, thus resulting in 22--60 nm deep wells for the following metal deposition. 20~nm of Ag with 2~nm of Ti as the adhesive layer was used for the metallization, thus resulting in the 22 nm thick metal layer (Fig.~\ref{Schematic}, grey layer). By omitting the final lift-off processing steps from the original DALI method, gapped bowtie antenna particles overlaid with a closely spaced metal film dotted with similarly gapped and aligned bowtie apertures were created.\cite{kabusure2023raman} Following fabrication, the structures were imaged using a scanning electron microscope (SEM) (Zeiss, Sigma VP) (Fig.~\ref{Schematic}c,d additional SEM images in Supporting Information Figs.~S1--S4).

\subsection*{Substrate preparation for Raman experiments}
Post fabrication, samples were spin-coated with a poly(methyl methacrylate) (PMMA) layer to reduce further oxidation of the Ag film. Before Raman measurements, the PMMA layer was first removed by immersing the substrates into acetone followed by sonication for $\sim$10 min. The samples were then cleaned with isopropanol (IPA) for $\sim$5 min to remove any residues and blow dried with a N$_2$ flow. A rhodamine 6G (R6G) solution was prepared as described previously\cite{kabusure2023raman} and spin-coated onto substrates using a spin coater at 3000 rpm for 30 s resulting in a $\sim$120~nm thick R6G-film.

\subsection*{Raman measurement}

Raman signals of R6G spin-coated on the structures were measured using a commercial Renishaw Invia Reflex Raman microscope with WiRE~$\textsuperscript{TM}$ software. The Raman equipment was first calibrated using the blank silicon wafer sample before taking measurements. The sample was placed under the microscope and focused using white light and a~50$\times$ objective lens. The area of interest was selected using a live feed camera in the software and microscope knobs before switching to the laser source to allow illumination of 785~nm laser excitation wavelength on the sample. To acquire spectrum, centre wavelength of 1500~cm$^{-1}$, accumulation of 1, and exposure time of 10 s were set. Raman spectra were averaged from 9 measurements covering an area of 50 $\times$ 50~$\mu$m$^2$ (3$\times$3 measurement grid with a 25 $\mu$m step size). The obtained Raman spectra were normalized by subtracting the smallest possible signal value from all data points in each spectrum.

\subsection*{Finite difference time domain (FDTD) simulations}

The electric field intensity enhancements of the aperture-particle hybrid structures were simulated using the finite-difference time-domain (FDTD) technique in Lumerical simulation software (Ansys). We used stabilized perfectly matched layers (PML) as simulation boundaries to minimize reflections, guaranteeing better simulation stability and more accurate results. The electric field of the incident light was set to 1~V~m$^{-1}$. The simulation waveband was chosen to be 450--1500~nm to cover potential Raman resonant wavelengths.

\section*{Results}

\begin{figure*}[t!]
    \centering
    \includegraphics[width=0.85\textwidth]{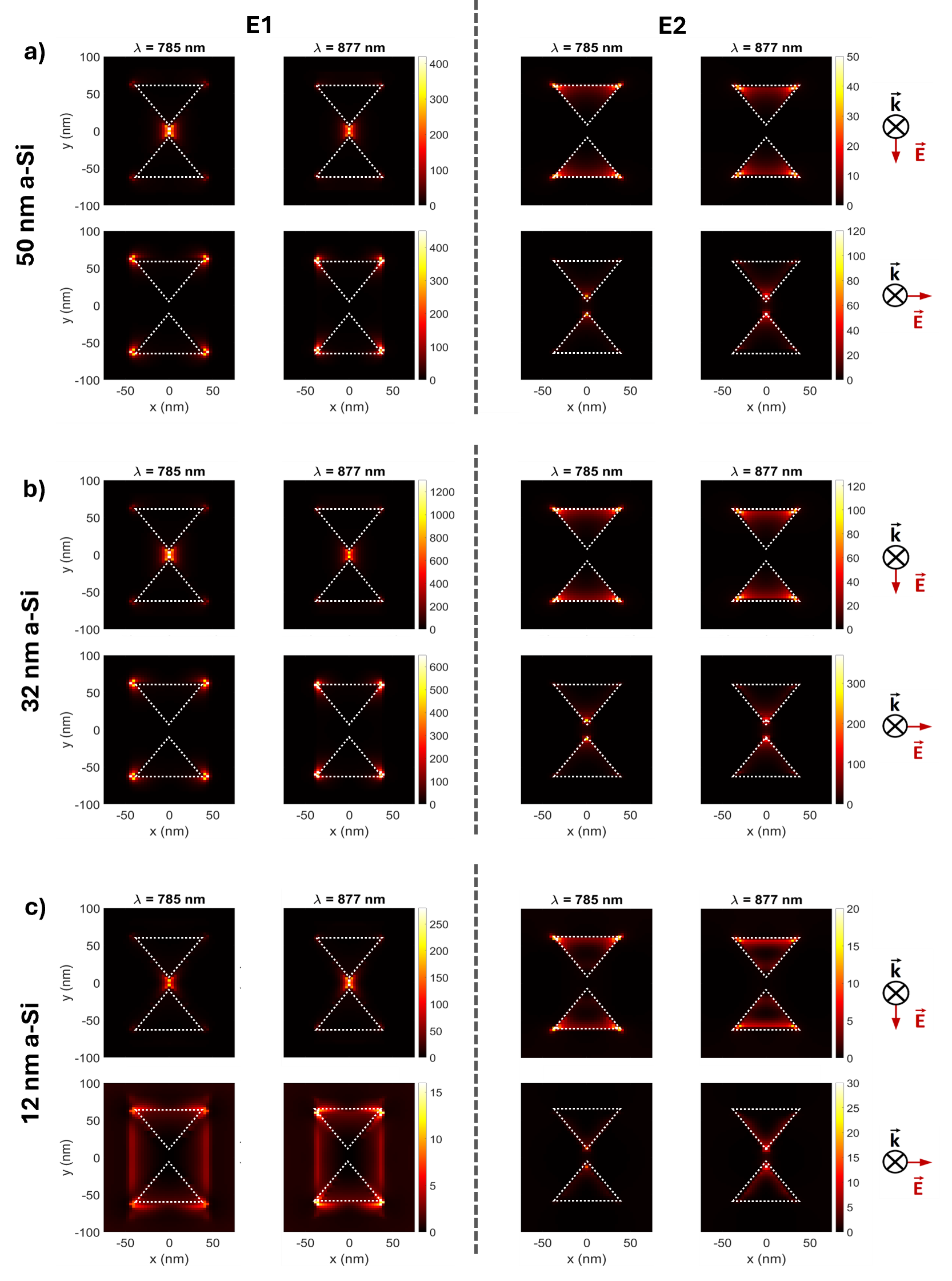}
    \caption{FDTD simulations showing electric field intensity profiles for samples with (a) 50~nm, (b) 32~nm, and (c) 12~nm a-Si spacer thickness at different polarization directions. E1 and E2 correspond to the planes indicated in Fig.~\ref{Schematic}. Top row shows the field enhancements with y-polarization and the bottom row with x-polarization. Left column corresponds to the wavelength of 785 nm (Raman laser wavelength) and the right column to 877 nm (the most prominent Raman transition of the R6G). The dashed lines indicate the outlines of the bowties and the apertures.}
  \label{FDTDE1E3}
\end{figure*}

In Fig.~\ref{FDTDE1E3} we show the FDTD simulated electric field enhancements for three different a-Si spacer layer thicknesses (50~nm, 32~nm and~12 nm; 42~nm sample is shown for simplicity in Supporting Information Fig.~S5). For each spacer layer thickness, we have plotted the enhancements at two planes of interest, namely at the midway of the bowtie particles and at the midway of the aperture layer (E1 and E2 planes indicated in Fig.~\ref{Schematic}). Additionally, we have presented the enhancements at two different wavelengths, 785~nm and 877~nm, which correspond to the wavelengths of the Raman laser and the most prominent Raman peak of R6G, respectively. The simulations were carried out using two different polarizations corresponding to the long ($y$) and short ($x$) axes of the bowties, respectively.

In Fig.~\ref{FDTDE1E3}a (50~nm a-Si), it can be seen that the enhancements were approximately 400 for both polarizations at the center plane of the bowtie particles (E1) and on the order of 50--120 at the plane of the apertures (E2). Thus one may conclude that in this case the overall field enhancement was primarily contributed by the particle resonances. Reducing the spacer layer thickness to 32~nm led to drastic changes in the field enhancements, as seen in Fig.~\ref{FDTDE1E3}b. The longitudinal ($y$-) polarization showed intensity enhancements on the order of 1200, and the $x$-polarization of approximately 600 at the plane of the particles (E1). Notably, also aperture layer (E2) enhancement for both polarizations had increased by a factor of 2.5. Thus, one can conclude that the interlayer coupling profoundly affects the overall enhancement of the hybrid structure. A natural question then is, whether additional reduction of the spacer layer thickness would result in even higher enhancements. In Fig.~\ref{FDTDE1E3}c we have plotted the enhancements for the 12 nm spacer layer. Perhaps counter-intuitively, the observed enhancements were the lowest of all three cases. Next, we attempt to clarify the reason for this behavior.

\begin{figure*}[ht]
\centering
  \includegraphics[width=\textwidth]{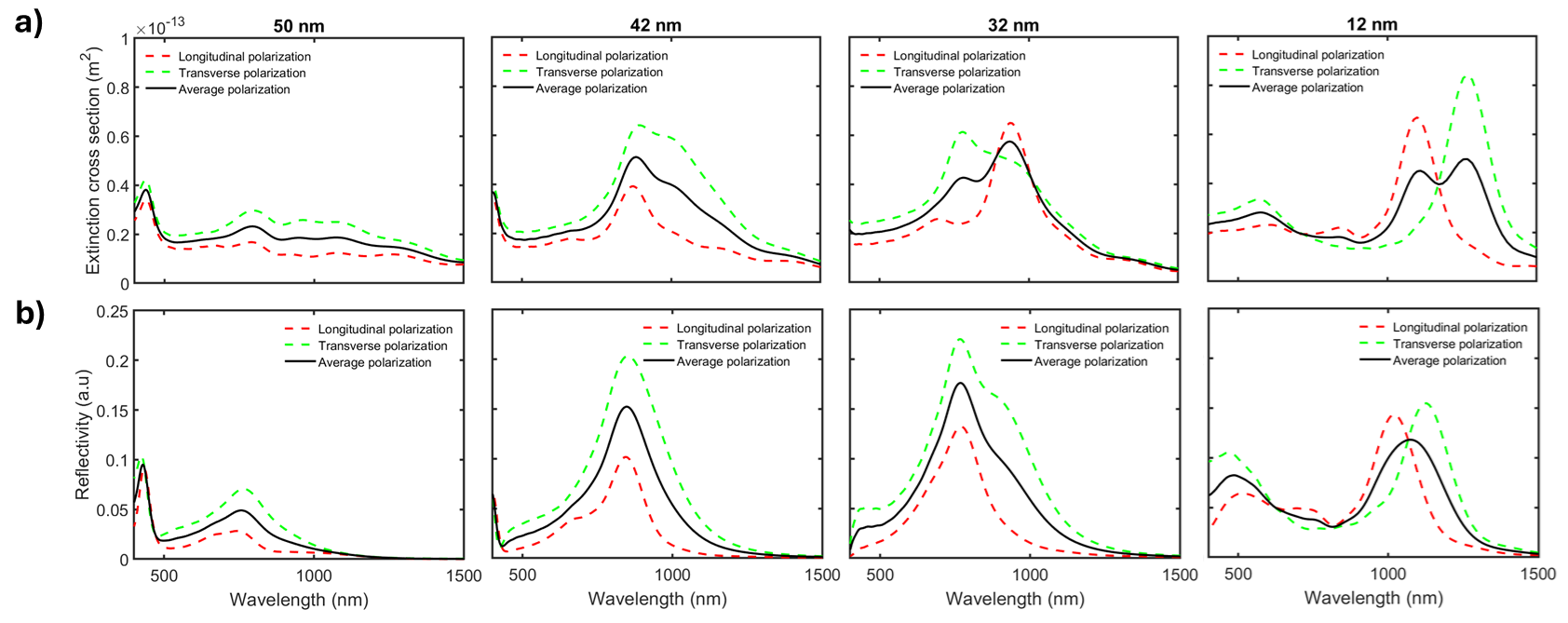}
 \caption{FDTD simulations showing both (a) extinction cross-section (m$^2$) and (b) reflectivity as a function of wavelength for a-Si spacer layers 50~nm, 42~nm, 32~nm, and 12~nm at different polarization directions.}
  \label{ExtinctionReflection}
\end{figure*}

In Fig.~\ref{ExtinctionReflection} we present the extinction cross-sections and the reflectivity as a function of the wavelength for the aforementioned structures (including 42~nm a-Si spacer layer thickness). The analysis was carried out for both longitudinal and transverse polarizations, and their average is shown as well. For the 50~nm a-Si layer, rather low values for both extinction and reflectivity were observed throughout the wavelength range. Reducing the layer thickness to 42~nm resulted in a significant increase in both quantities, with a clear resonant peak in the wavelength range of 750--1000 nm. For the 32-nm thick layer, the overall values for both extinction and reflectivity were similar to 42~nm case, but clear spectral changes could nevertheless be observed due to the increased interlayer coupling. Importantly, while the extinction and reflectivity were still fairly large for the 12~nm a-Si thickness, the spectral positions of the maxima had shifted to a wavelength range that no longer overlaps with the Raman laser (785 nm) or the Raman transitions of R6G (873-890 nm). Such wavelength shifts are typical in various systems, where two or more resonances couple strongly with each other. \cite{hakala2009vacuum,torma2015strong} We thus conclude that the low field enhancement values for the 12~nm spacer layer observed in Fig.~\ref{FDTDE1E3}c occur due to strong (also known as normal mode) coupling between the particle and aperture resonances, which shifts the resonances away from the wavelength range of the Raman laser and transitions.

\begin{figure}[ht]
    \centering
    \includegraphics[width=\columnwidth]{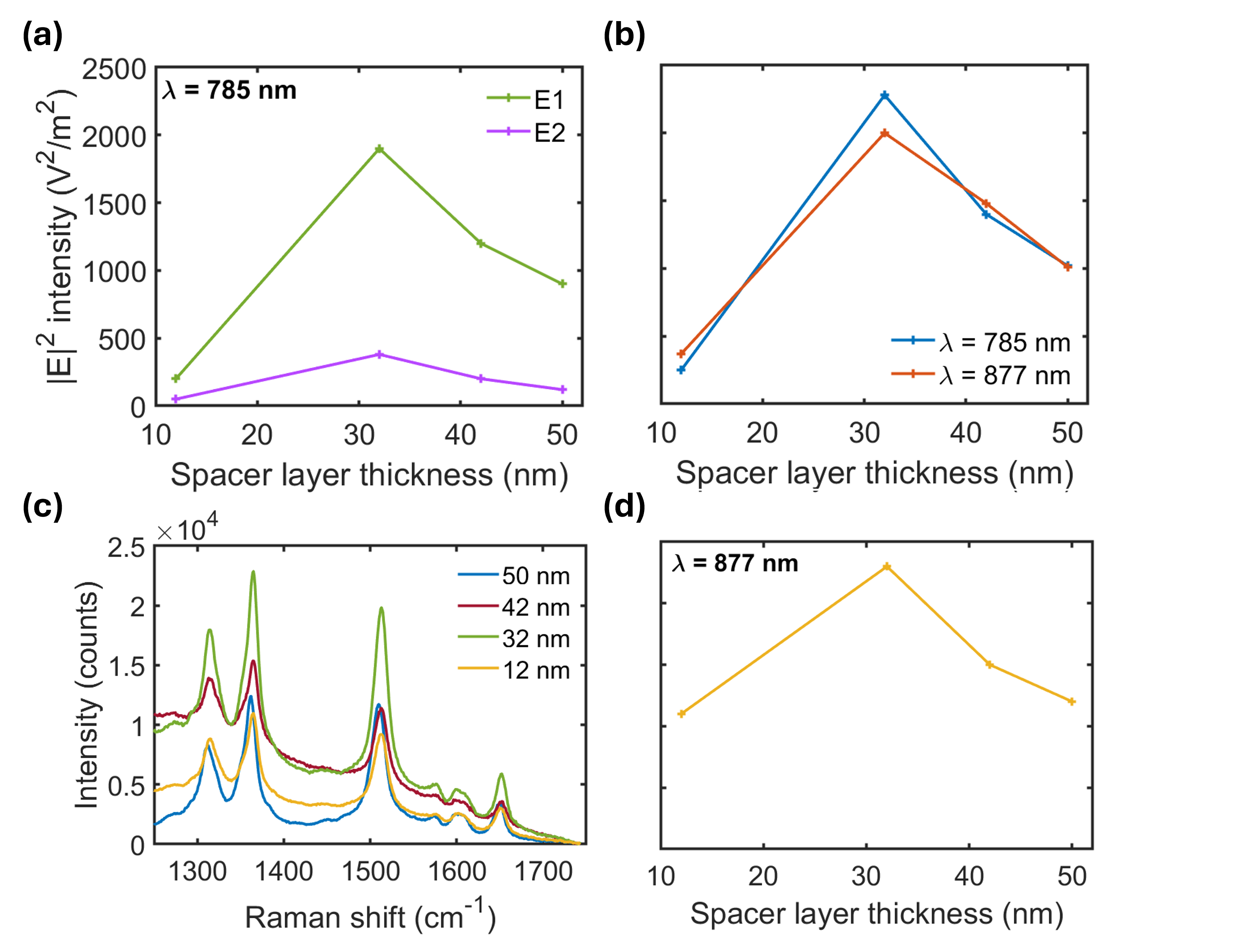}
    \caption{(a) The FDTD simulated electric field intensity maxima (sum of the maxima from two polarization directions) (V$^2$~m$^{-2}$) at E1 (green) and at E2 (purple) at the Raman excitation wavelength (785~nm). (b) The sum of the FDTD simulated electric field intensity maxima at E1 and E2 (V$^2$~m$^{-2}$) at the Raman excitation wavelength 785~nm (blue) and at 877~nm (red). (c) Measured Raman spectra of the particle-aperture hybrid structures with interlayer spacings varying from 50~nm to 12~nm. (d) Measured Raman intensities at 877~nm as a function of a-Si spacer layer thickness.}
    \label{excelfile}
\end{figure}

In Fig.~\ref{excelfile} we summarize the findings from the FDTD simulations and compare them with the experimentally obtained Raman signals from the samples with four different spacer layer thicknesses. Fig.~\ref{excelfile}a shows electric field intensities at the Raman laser wavelength at specific positions of the sample. We chose the positions that provided the maximum field enhancement based on Fig.~\ref{FDTDE1E3} results. For longitudinal polarization and the bowtie particles (plane E1), this position lies right in the gap of the bowtie (see Fig.~\ref{FDTDE1E3}a), top left corner). For transverse polarization, the other four remaining corners of the bowtie provided the highest enhancement. For apertures (plane E2), the four outermost corners gave the highest enhancement for the longitudinal polarization, and for the transverse polarization, the corners closest to each other had the highest enhancement, (see Fig.~\ref{FDTDE1E3}2a, right column).
Summing the enhancements for both polarizations thus gives an approximate measure of the overall enhancement at Raman laser wavelength, as seen in Fig.~\ref{excelfile}a. From this, it could be observed that, indeed, the 32-nm spacer layer provided the highest enhancement for both the particles and apertures. In Fig.~\ref{excelfile}b, we then show the sum of the intensities from both particle (E1) and aperture layers (E2). In addition, we noticed that the overall behaviour is very similar for both the Raman laser and the Raman transition wavelengths (785 and 877~nm).

Finally, the experimentally obtained Raman intensities from the four R6G coated samples are presented in Fig.~\ref{excelfile}c. Notably, the 12~nm a-Si thickness resulted in the lowest overall Raman signal. The highest signal, on the other hand, was recorded from the sample with a 32~nm thick spacer layer. The obtained Raman intensities as a function of spacer layer thickness at the wavelength of 877 nm (which corresponds to the most prominent Raman peak in Fig.~\ref{excelfile}c) have been plotted in Fig.~\ref{excelfile}d. The order of samples with increasing Raman signal was as follows: 12, 50, 42 and 32 nm a-Si thickness. Comparing Figs.~\ref{excelfile}b and \ref{excelfile}d, we observed a clear correlation between the simulated field enhancements and the measured Raman signals.

While reducing the spacer layer thickness results in a stronger interlayer coupling, a too thin spacer layer induces resonance wavelength shifts that lie outside the spectral window of the Raman laser and the Raman transitions. We thus conclude that there exists an optimal spacing for a selected layer conformation/geometry that gives the highest field enhancement and the strongest Raman signal, which in our geometry is achieved using a 32~nm spacer layer thickness.

\section*{Conclusions}

To conclude, we have presented a DNA-assisted lithography fabricated SERS substrate consisting of two-layer hybrid nanostructures. One layer of the substrate included bowtie-shaped silver nanoparticles and the other similarly shaped and aligned apertures in a silver film. The effects of interlayer spacing revealed that there is significant coupling between the layers, which further enhanced the optical near fields as implied by the FDTD simulations and the experimentally observed Raman signals from R6G-coated substrates. Since the aperture film is generated automatically during lithographic processing, the approach may provide remarkable advantages over common single-layer Raman substrates, without additional fabrication cost. In our specific case, we found that while the reduction of the interlayer spacing may increase the near field enhancements, a too thin spacer results in too strong spectral modifications of the resonances, such that they shift beyond the wavelength range of interest for both the Raman excitation laser and the Raman transitions. A natural follow-up work would include \textit{a-priori} design of the structure, so that the normal mode coupling would result in both the maximum field enhancement and the resonance wavelengths that overlap with the Raman excitation as well as the transitions of the molecule of interest.

\section*{Supporting Information Available}

  Additional SEM images of the samples, additional FDTD simulations (PDF)

\begin{acknowledgements}
The authors thank the Research Council of Finland  (359450, 339544, PREIN Flagship 346518), the Emil Aaltonen Foundation, the Sigrid Jusélius Foundation, the Jane and Aatos Erkko Foundation, the Magnus Ehrnrooth Foundation, the Finnish Cultural Foundation (Kalle and Dagmar Välimaa Fund), ERA Chair MATTER from the European Union's Horizon 2020 Research and Innovation Programme under Grant Agreement No.~856705, and Mobilitas 3.0 Programme under the framework of the JPIAMR - Joint Programming Initiative on Antimicrobial Resistance. We also acknowledge the provision of facilities and technical support by Aalto University Bioeconomy Facilities, OtaNano – Nanomicroscopy Center (Aalto-NMC), and Micronova Nanofabrication Center.
\end{acknowledgements}

\begin{contributions}
Kabusure M. Kabusure: 0000-0003-4839-1189 \\
Petteri Piskunen: 0000-0002-3142-3191 \\
Jarkko J. Saarinen: 0000-0002-0537-0452 \\
Veikko Linko: 0000-0003-2762-1555 \\
Tommi K. Hakala: 0000-0003-3853-4668  
\end{contributions}

\begin{interests}
The authors declare no competing financial interest.  
\end{interests}

\end{document}


\title{\LARGE{Supporting Information} \\ \vspace{0.7 cm} Controlling Raman enhancement in particle-aperture hybrid nanostructures by interlayer spacing}
\shorttitle{Supporting Information}

\author[1,2$\dagger$]{Kabusure M. Kabusure}
\author[3,$\dagger$]{Petteri Piskunen}
\author[2]{Jarkko J. Saarinen}
\author[3,4,*]{Veikko Linko}
\author[1,*]{Tommi K. Hakala}

\affil[1]{Department of Physics and Mathematics, University of Eastern Finland, Yliopistokatu 2, P.O Box 111, 80101, Joensuu, Finland}
\affil[2]{Department of Chemistry, University of Eastern Finland, Yliopistokatu 7, P.O Box 111,
FI-80101, Joensuu, Finland}
\affil[3]{Biohybrid Materials, Department of Bioproducts and Biosystems, Aalto University, P.O. Box 16100, 00076 Aalto, Finland}
\affil[4]{Institute of Technology, University of Tartu, Nooruse 1, 50411, Tartu, Estonia}
\affil[$\dagger$]{Equal contribution}
\affil[*]{Correspondence and requests for materials should
be addressed to \href{mailto:veikko.pentti.linko@ut.ee}{veikko.pentti.linko@ut.ee} or
\href{mailto:tommi.hakala@uef.fi}{tommi.hakala@uef.fi}}
\maketitle

\onecolumn

\clearpage
\section{Supplementary SEM images}

\begin{figure}[h!]
    \centering
    \includegraphics[width=0.9\columnwidth]{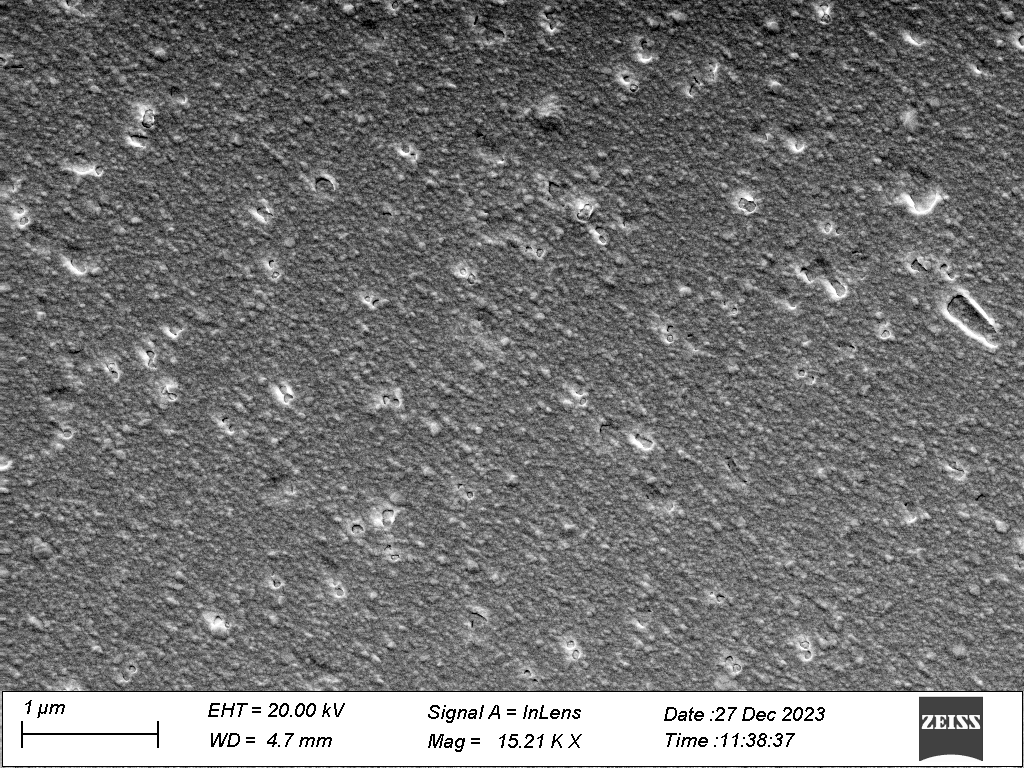}
    \caption{Large-scale SEM image of the particle-aperture patterns with a 12~nm spacer layer.}
    \label{Fig_S1}
\end{figure}

\begin{figure}[h!]
    \centering
    \includegraphics[width=0.9\columnwidth]{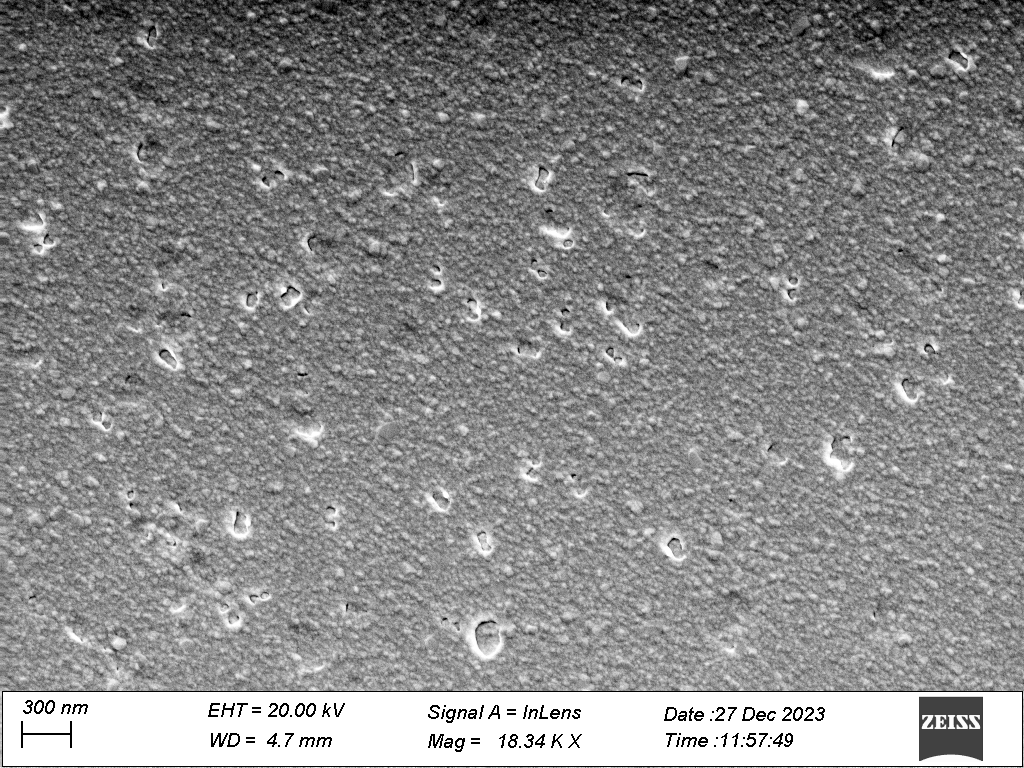}
    \caption{Large-scale SEM image of the particle-aperture patterns with a 12~nm spacer layer.}
    \label{Fig_S2}
\end{figure}

\begin{figure}[h!]
    \centering
    \includegraphics[width=0.9\columnwidth]{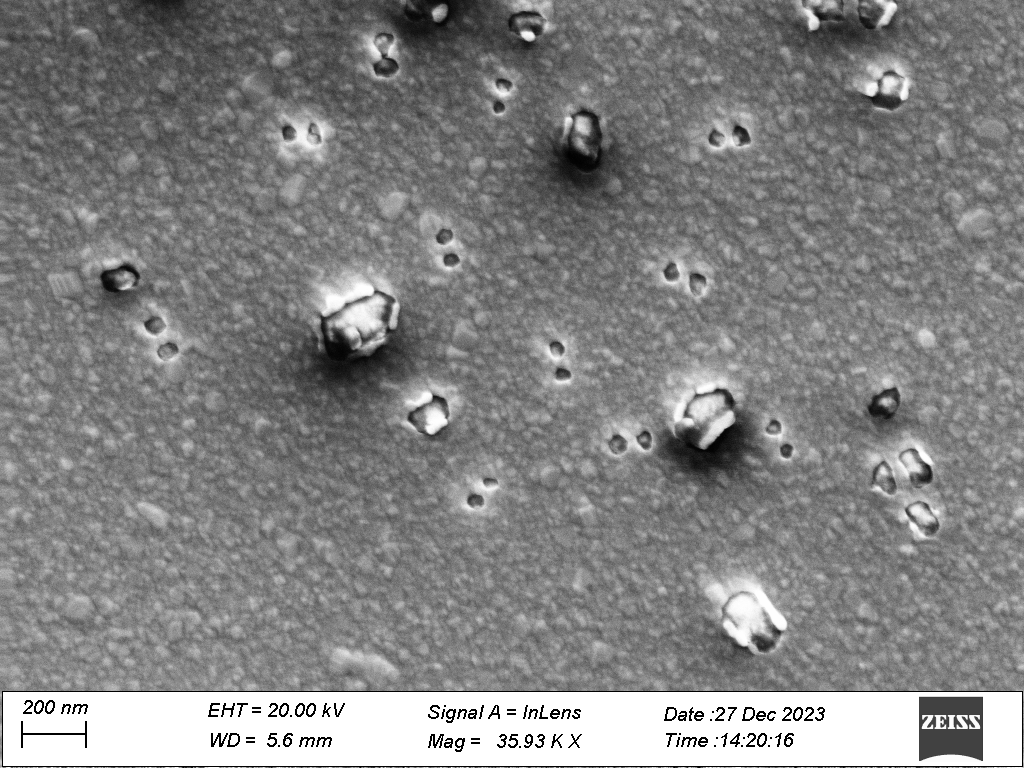}
    \caption{Large-scale SEM image of the particle-aperture patterns with a 50~nm spacer layer.}
    \label{Fig_S3}
\end{figure}

\begin{figure}[h!]
    \centering
    \includegraphics[width=0.9\columnwidth]{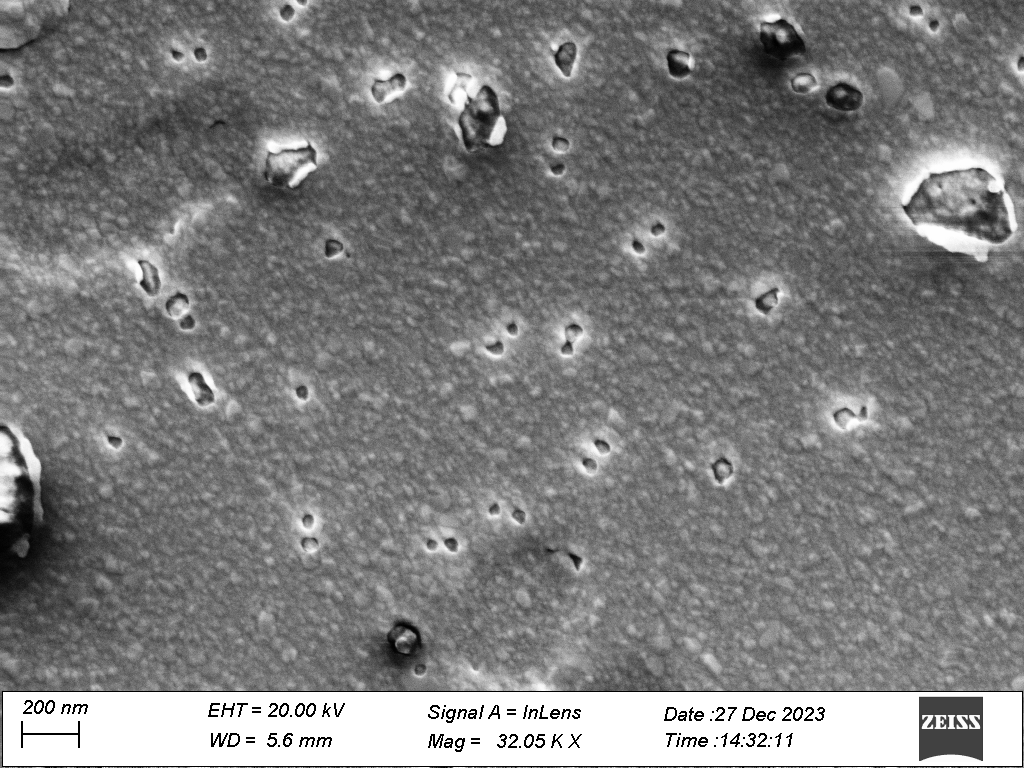}
    \caption{Large-scale SEM image of the particle-aperture patterns with a 50~nm spacer layer.}
    \label{Fig_S4}
\end{figure}

Figures~\ref{Fig_S1}--~\ref{Fig_S4} show representative large-scale SEM images of the Ag particle-aperture patterns at 12~nm and 50~nm spacer thicknesses.

\clearpage

\section{Additional FDTD simulations}

\begin{figure}[h!]
    \centering
    \includegraphics[width=0.9\columnwidth]{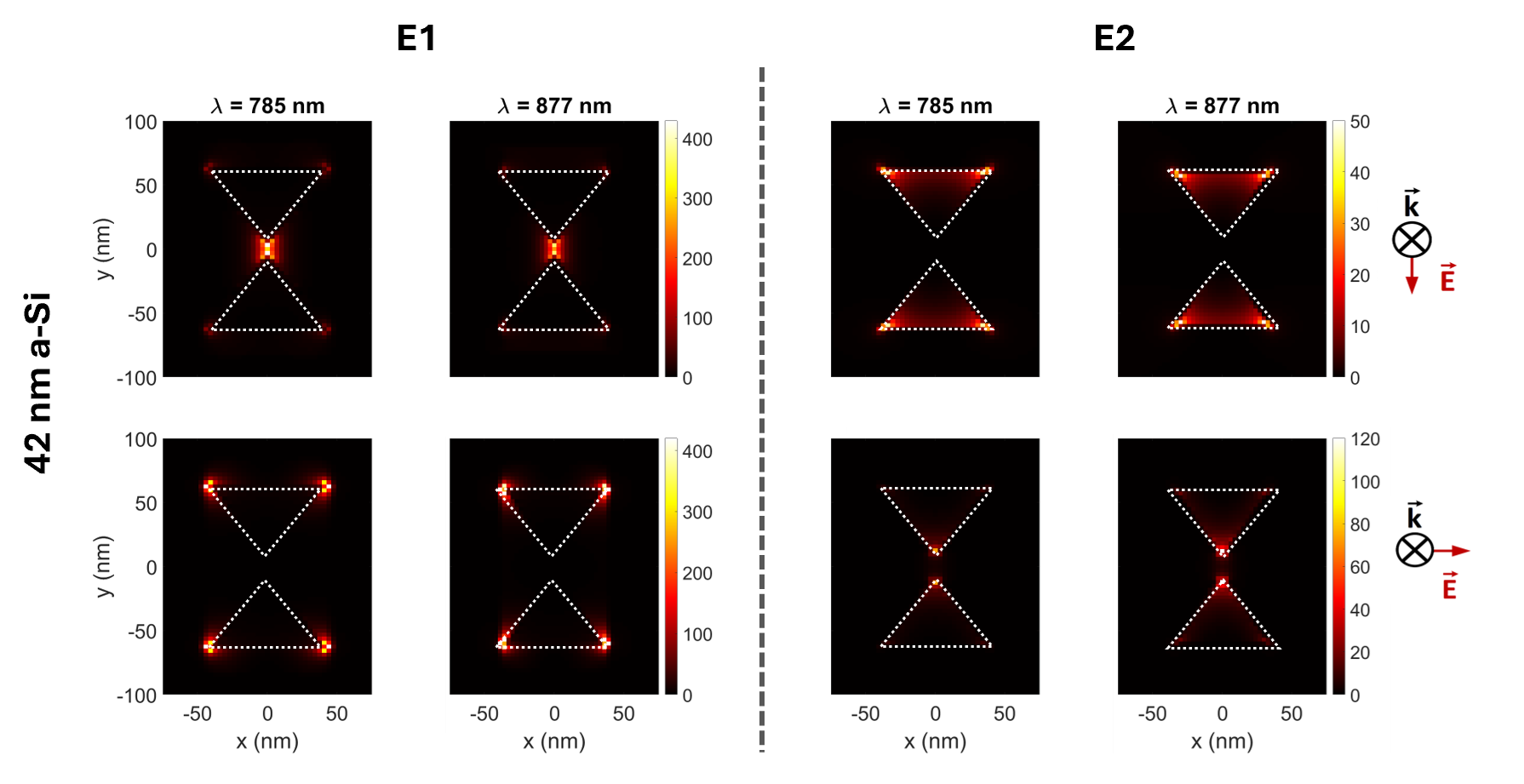}
    \caption{FDTD simulations showing electric field intensity profiles for a sample with 42~nm a-Si spacer thickness at different polarization directions. E1 and E2 correspond to the planes indicated in main article Fig.~1. Top row shows the field enhancements with y-polarization and the bottom row with x-polarization. Left column corresponds to the wavelength of 785 nm (Raman laser wavelength) and the right column to 877 nm (the most prominent Raman transition of the R6G). The dashed lines indicate the outlines of the bowties and the apertures.}
    \label{Fig_S5}
\end{figure}